\documentclass[%
 reprint,
superscriptaddress,
 amsmath,amssymb,
 aps,prl
]{revtex4-2}

\usepackage{graphicx}
\usepackage{dcolumn}
\usepackage{bm}

\begin{document}

\preprint{APS/123-QED}

\title{Electronic state back action on mechanical motion \\in a quantum point contact coupled to a nanomechanical resonator}

\author{Andrey A. Shevyrin}
 \email{shevandrey@isp.nsc.ru}
\affiliation{%
 Rzhanov Institute of Semiconductor Physics SB RAS\\
13 Lavrentyeva ave., Novosibirsk, Russia, 630090
}%

\author{Askhat K. Bakarov}
\affiliation{%
 Rzhanov Institute of Semiconductor Physics SB RAS\\
13 Lavrentyeva ave., Novosibirsk, Russia, 630090
}%
\affiliation{%
 Novosibirsk State University\\
2 Pirogova str., Novosibirsk, Russia, 630090
}
\author{Alexander A. Shklyaev}
\affiliation{%
 Rzhanov Institute of Semiconductor Physics SB RAS\\
13 Lavrentyeva ave., Novosibirsk, Russia, 630090
}
\affiliation{%
 Novosibirsk State University\\
2 Pirogova str., Novosibirsk, Russia, 630090
}
\author{Arthur G. Pogosov}
\affiliation{%
 Rzhanov Institute of Semiconductor Physics SB RAS\\
13 Lavrentyeva ave., Novosibirsk, Russia, 630090
}
\affiliation{%
 Novosibirsk State University\\
2 Pirogova str., Novosibirsk, Russia, 630090
}


\date{April 8, 2024}

\begin{abstract}
In a nanomechanical resonator coupled to a quantum point contact, the back action of the electronic state on mechanical motion is studied. The quantum point contact conductance changing with subband index and the eigenfrequency of the resonator are found to correlate. A model is constructed explaining the frequency deviations by the alternating ability of the quantum point contact to screen the piezoelectric charge induced by mechanical oscillations. The observed effects can be used to develop electromechanical methods for studying the density of states in quasi-one-dimensional systems.
\end{abstract}

\maketitle


Due to advances in nanofabrication, mechanical resonators having sub-micron dimensions \cite{Schmid_2023} and ultra-small mass can exhibit high quality factors and resonant frequencies \cite{Engelsen_2024_dilution}. This unique combination of properties makes nanomechanical resonators interesting as artificial systems capable of displaying the transition between classical and quantum motion \cite{Samanta_2023_ground,Poot_2012} and complex nonlinear dynamics \cite{Huber_2020_nonlinear,Bachtold_2002_review}. In addition, nanomechanical systems are promising for sensing applications and quantum computing \cite{Pistolesi_2021_qubits}.

The devices based on nanomechanical resonators, apart from mechanical degrees of freedom, can have additional electrical functionality. Electromechanical coupling of electrostatic, magnetoelectric or piezoelectric nature can be used for on-chip driving and detection of nanomechanical motion \cite{Ekinci_2005,Schmid_2023}. Most importantly, when system dimensions are reduced, this coupling gives birth to new effects unobservable in macroworld.

Downscaling of electron systems is well known to change the laws of electron transport \cite{Ihn_2009,Datta_1995}, making the ballistic and tunnel regimes accessible, and to open up a great variety of quantum effects, ranging from conductance quantization and the Coulomb blockade to the actively studied many-body effects \cite{Hew_2008_interaction,Kumar_2014_interac}. The question of how these drastic changes affect the electromechanical coupling phenomena remains largely open \cite{Yamaguchi_2017_review,Pogosov_2022_review}.

The direct action of mechanical motion on electron transport has been revealed in 2-,1- and 0-dimensional systems \cite{Shevyrin_2016_piezo,Steele_2009_QD,Cleland_2002,Okazaki_2013_QPC,Poggio_2008_QPC,Pashkin_2010_SET}. A two-dimensional electron gas (2DEG), quantum point contacts (QPCs), quantum dots (QDs) and single-electron transistors have been shown to be sensitive enough to mechanical motion to discuss their use even when approaching the quantum ground state of nanomechanical resonators \cite{Knobel_2003_SET}.

This naturally raises the importance of the back action of quantum transport on mechanical motion, which, although currently perhaps even more interesting than the direct one, remains less studied. Such back action is used as a way of cooling and amplification of motion, and its possible applications for electrically controlled phonon lasing and single-phonon emission are considered \cite{Naik_2006_cooling,Okazaki_2016_QD}. Apart from that, the back action can be used to study electron transport within electrically isolated systems, such as double QDs \cite{Khivrich_2019_DQD}, converting the inter-dot electron tunneling into a change in the most precisely measurable physical quantity, namely, the frequency of oscillations.

Compared to 2DEG \cite{Yamaguchi_2012_2DEG} and QDs \cite{Steele_2009_QD,Okazaki_2016_QD}, the electromechanical back action in quasi-1D systems is poorly studied \cite{Mozyrsky_2002,Benatov_2012,Clerk_2004,Stettenheim_2010}. In QPCs, its signatures have been revealed in the spectrum of shot noise in the tunneling regime \cite{Stettenheim_2010}, close to the pinch-off. An important question of whether dispersive and dissipative back action can be observed in the ballistic regime, when the oscillation period far exceeds the time an electron spends in a QPC, remains open. 

Such back action may give an additional experimental method for nanomechanical probing of the density of states (DOS) and thus contribute to the understanding of the many-body effects observed in QPCs. The measurement of compressibility \cite{Smith_2011_compressibility,Luscher_2007_compressibility,CASTLETON_1998_compressibility}  and quantum capacitance \cite{Drexler_1994_qcapacitance,Ilani_2006_qcapacitance,Jarrat_2020_qcapacitance} in QPCs, also giving important information about the DOS, is known to be challenging, because, in such experiments, the measured signal is mostly determined by the parasitic and geometric, rather than quantum, capacitances. Thus, the back action in quasi-1D systems is promising for both nanoelectromechanics and quantum transport research.

In the present manuscript,  we report on experimental observation of a dispersive back action in a nanomechanical resonator coupled to a QPC, which is in the ballistic regime of electron transport, and show that the observed resonant frequency deviations can be related to the quasi-1D density of states via the piezoelectric effect.

\begin{figure*}
    \includegraphics{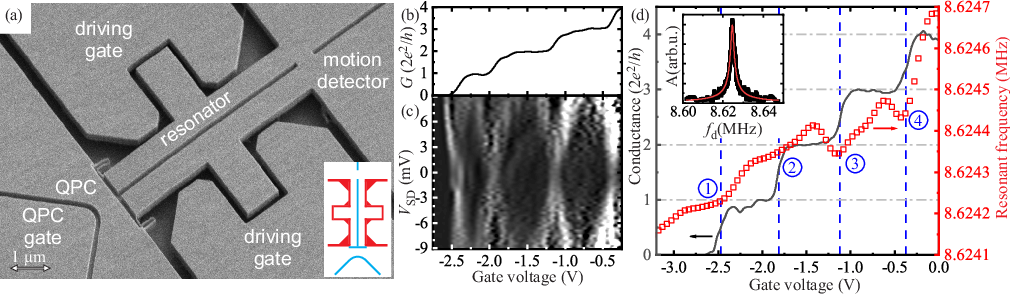}
    \caption{\label{fig:fig1}(a) An SEM image of a cross-shaped doubly clamped nanomechanical resonator coupled to a quantum point contact (QPC). Inset: shallow (light blue) and deep (red) etching is schematically shown (adapted from \cite{Shevyrin_2021}). (b) Conductance $G$ of the QPC as a function of the voltage $V_\mathrm{G}$ applied to the electrode marked "QPC gate" and the resonator playing the role of a second gate. (c) $\partial{G}/\partial V_\mathrm{G}$ measured as a function of $V_\mathrm{G}$ and source-drain voltage $V_{SD}$. The black diamonds are the conductance plateaus. (d) Eigenfrequency of the resonator (red squares) correlating with the QPC conductance (black solid line). Inset: the dependence of the signal $A$ proportional to the vibration amplitude on the driving frequency $f_\mathrm{d}$.}
\end{figure*}

Experimental samples are created from the AlGaAs/GaAs heterostructure described in details elsewhere \cite{Shevyrin_2014_QPC}. The heterostructure comprises a 2DEG at the depth of 88 nm under the surface. An image of one of the samples is shown in Fig. \ref{fig:fig1} (a). The sample is a nanomechanical resonator shaped as a cross-like bridge, 166 nm thick, 6 $\mu$m long and 1.3 $\mu m$ wide. Near one of its clamped edges, a QPC is placed, separated from the resonator and a nearby gate ("QPC gate") by shallow (80 nm deep) and narrow (150 nm wide) plasmachemically etched trenches. At another clamped edge of the resonator, a 2DEG constriction is located which is used as a piezoresistive detector of mechanical motion ("motion detector"). The resonator is surrounded by two coplanar side gates ("driving gate") used to drive its mechanical motion as described below.

First, the QPC conductance $G$ has been measured as a function of the gate $V_\mathrm{G}$ and source-drain $V_\mathrm{SD}$ voltages to characterize its electron-transport properties, with the resonator being immobile. After that, the resonator motion was driven, and the influence of the QPC state on mechanical and electromechanical properties of the system has been studied. All the measurements has been done at $T=$1.6 K.

In the first set of measurements, the gate voltage was applied to the "QPC gate" (see Fig. \ref{fig:fig1} (a)) and to the resonator playing the role of a second side gate, while the QPC potential remained close to zero. The conductance was measured using lock-in technique at the frequency of 77 Hz and the amplitude of 30 $\mathrm{\mu V}$. In Fig. \ref{fig:fig1} (b), $G(V_\mathrm{G})$ dependence displaying the plateaus is shown (the resistance of the contacts and mesa, close to 10 $\mathrm{k\Omega}$, is subtracted). The conductance derivative with respect to $V_\mathrm{G}$ is shown in the color plot in Fig. \ref{fig:fig1} (c) as a function of $V_\mathrm{G}$ and $V_\mathrm{SD}$ (dc component), where the black diamond-like regions are the quantization plateaus, and their vertical dimensions can be used to extract the inter-subband energy separations \cite{Glazman_1989_nonlin}. With the contact resistance taken into account, this gives the inter-subband separations 4, 2.8 and 2.2 meV for the first, second and third diamonds, respectively.

After electrical characterization of the QPC, its source and drain terminals were both connected to a dc voltage source, while the potentials of the resonator and the QPC gate remained close to zero. The QPC state is determined by the potential difference between the QPC and the surrounding electrodes, and applying positive voltage $-V_\mathrm{G}$ to the QPC in the latter configuration is identical to applying the opposite negative voltage $V_\mathrm{G}$ in the first set of measurements, which makes it possible to compare the dependences on $V_\mathrm{G}$. Zero voltage between the QPC source and drain eliminates the need to consider the possible role of heating effects in the observed phenomena.

Mechanical oscillations of the resonator were driven electrostatically at its fundamental frequency corresponding to out-of-plane flexural vibrations \cite{Shevyrin_2015_actuation} by applying a sum of dc (4 V) and ac (50 mV RMS) voltages to the driving gates. To detect the oscillations, we measured the conductance response of the 2DEG constriction labeled as "motion detector" in Fig. \ref{fig:fig1} (a) using on-chip heterodyne down-mixing \cite{Bargatin_2005_downmixing,Shevyrin_2016_piezo}. The inset in Fig. \ref{fig:fig1} (d) shows a typical dependence of the measured signal (current) on the driving frequency, which has a lorentzian shape. The resonant frequency $f$ is close to 8.62 MHz, and the quality factor is about 2600.

At each $V_\mathrm{G}$ value, the resonant frequency $f$ has been derived by fitting such dependences on the driving frequency. The resulting $f(V_\mathrm{G})$ dependence is shown in Fig. \ref{fig:fig1} (d), together with $G(V_\mathrm{G})$ dependence. The displayed curve, on a smooth background (increasing with $V_\mathrm{G}$ increase), demonstrates oscillations clearly correlated with the QPC conductance: $f$ displays local mimima each time when $G$ shows the inter-plateau risers (labeled in the figure). The frequency minima observed at the first and second inter-plateau risers are smooth, and the second minimum is shallower than the first one. The third and the fourth minima are deeper, and, because of their abruptness, they look qualitatively different from the first two. The frequency oscillations are small but robust and qualitatively reproducible.

The dependence of the resonant frequency on $V_\mathrm{G}$ shows that the effective stiffness of the system is affected by its electronic state, while the correlation between the frequency and the QPC conductance indicates an important role of the QPC in this back action. The effect can be described in terms of the potential energy of the system. Let $\xi$ be the displacement of the central point of the resonator. Then the potential energy is $W=\frac{m^*(2\pi f_0)^2\xi^2}{2}+U(\xi)$, where the first term is the mechanical energy, the second term is the sum of electrical and chemical Gibbs energies depending on $\xi$, and $m^*$ is the effective mass of the resonator. The deviation $f-f_0$ of the resonant frequency can be calculated as
\begin{equation}
    \Delta f=\frac{1}{8\pi^2 m^*f_0}\frac{\mathrm{d}^2U}{\mathrm{d}\xi^2}
\end{equation}
(at $\lvert\Delta f\rvert\ll f_0$). Below we discuss two possible physical mechanisms that may cause $U$ dependence on $\xi$, and, thus, $f$ dependence on $V_\mathrm{G}$, namely, the capacitive (electrostatic) and piezoelectric coupling.

The effect of electrostatic softening/hardening \cite{Wu_2011_soft_hard,Shevyrin_2015_actuation}, typically observed when voltage $V_\mathrm{G}$ is applied between a resonator and an electrode nearby (QPC in our case) is caused by the oscillation-induced variation of capacitance $C$ between them. The corresponding frequency shift is
\begin{equation}
    \Delta f_\mathrm{es}=-\frac{1}{16\pi^2 m^*f_0}\frac{\mathrm{d}^2C}{\mathrm{d}\xi^2}(V_\mathrm{G}-V_\mathrm{G0})^2,
\end{equation}
where $V_\mathrm{G0}$ is the "charge neutrality" point. Generally speaking, as the QPC has a finite DOS, $C$ depends on its electronic state because of the quantum capacitance contribution \cite{Luryi_1988_qcapacit,Blien_2020_qcapac,Kim_2019_qcapac} ($C^{-1}=C^{-1}_\mathrm{geom}+C^{-1}_\mathrm{q}$, where $C_\mathrm{geom}$ and $C_\mathrm{q}$ are the geometric and quantum capacitances, respectively). The ratio $C_\mathrm{geom}/C_\mathrm{q}$ is of the order of the ratio of the vertical and horizontal dimensions of the diamonds in Fig. \ref{fig:fig1} (c), i.e $C_\mathrm{geom}/C_\mathrm{q}\approx0.004$. A change $\delta C_\mathrm{q}$ in the quantum capacitance would lead to a relative change in coefficient $\frac{\mathrm{d}^2C}{\mathrm{d}\xi^2}$ of the order of $C_\mathrm{geom}\delta C_\mathrm{q}/C_\mathrm{q}^2\ll 1$. Thus, the capacitive frequency shift would display only weak oscillations due to the DOS change on a much stronger parabolic background of $f(V_\mathrm{G})$ dependence, and it can hardly explain the observed frequency oscillations. Since the motion amplitude tends to zero at the clamped edge, the resulting capacitance change should be also vanishingly small, although it may contribute to the smooth background of $f(V_\mathrm{G})$ dependence.

Let's discuss the piezoelectric coupling, which, as we believe, is responsible for the frequency oscillations in our case, and whose role is enhanced by the QPC placement at the clamped edge, where the mechanical stress induced by the resonator oscillation is maximal. Likewise the sound velocity in a material is affected (namely, increased) by its piezoelectric properties, we can expect the elastoelectric coupling to modify (namely, increase) the resonator eigenfrequencies.  The stress induces an alternating piezoelectric bound charge $\rho_\mathrm{piezo}(\vec{r})\propto \xi$ in the bulk and at the surface \cite{Shevyrin_2016_piezo}. The QPC partially screens the piezoelectric field, adjusting the spatial distribution of the electron density, and reduces the magnitude of the considered effect, lowering the system energy. The screening effectiveness is determined by the DOS in the QPC which oscillates with subbands filling. The resonant frequency minima appear at the conductance inter-plateau transitions, where the DOS and the screening effectiveness are maximal, and this speaks in favour of the proposed explanation based on the piezoelectric effect.

However, the QPC, being an adiabatic 2DEG constriction, cannot be characterized by a single DOS, because it is coordinate-dependent. Explicit modelling of the effect requiring 3D simulation of mechanical stress and bound charge $\rho_\mathrm{piezo}(\vec{r})$ for the real geometry, as well as Schr$\mathrm{\ddot{o}}$dinger-Poisson calculations, go beyond the framework of the present paper. Instead, we limit ourselves to a model minimally needed to qualitatively reproduce the correlation between the resonant frequency and the QPC conductance under the conditions when the DOS depends on single coordinate $x$, and the temperature is finite.

The only direct consequence of the mechanical oscillations that we put in our model is the appearance of piezoelectric charge $\rho_\mathrm{piezo}(\xi,\vec r)=\rho_0(\xi)\chi(\vec r)$. The screening charge is calculated using the Thomas-Fermi approximation in the assumption of quasi-1D thermally smoothed DOS $\tilde D(x,V_\mathrm{G})$ \cite{Buttiker_1990}. The geometry is reduced to a symmetric one with respect to $x$-axis, with $t$ and $h$ being the transverse dimensions of the QPC and of the region occupied by $\rho_\mathrm{piezo}$, respectively. The $V_\mathrm{G}$-dependent part of the resonant frequency change can be evaluated via the energy of the system as
\begin{equation}\label{eq1}
    \Delta f_\mathrm{piezo}=\frac{e^2}{8\pi^2 f_0 m^*}\left(\frac{\partial\rho_0}{\partial \xi}\right)^2\Delta\Phi,
\end{equation}
where
\begin{equation}\label{dPhi}
\Delta\Phi=-\int\tilde D(x,V_\mathrm{G})\frac{\partial\phi}{\partial\rho_0}\frac{\partial\phi_\mathrm{piezo}}{\partial\rho_0}\,\mathrm{d}x,
\end{equation}
is the only $V_\mathrm{G}$-dependent multiplier in Eq. \ref{eq1}, $\phi_\mathrm{piezo}$ is the electric potential created by $\rho_\mathrm{piezo}$ only and $\phi$ is the total electric potential (including the contribution from the QPC charge). Hereinafter we will refer to $\Delta\Phi$ as to the resonant frequency change expressed in arbitrary units. Details of the calculation can be found in the Appendix.

\begin{figure}
\includegraphics{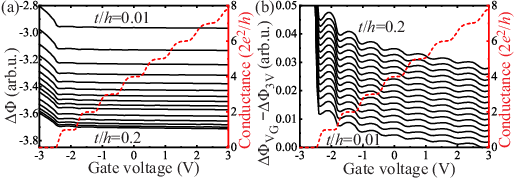}
\caption{\label{fig:fig2}(a) The calculated resonant frequency shift $\Delta \Phi$ (solid black curves) caused by the piezoelectric effect, with the quantum point contact (QPC) screening the piezoelectric field. Different curves correspond to different transverse dimensions $t$ of the QPC ($t/h$ is varied from 0.01 to 0.1 with 0.01 step and from 0.12 to 0.2 with 0.02 step, $h=166$ nm is the transverse size of the region occupied by the piezoelectric charge induced by mechanical oscillations). The dashed red line shows the QPC conductance. (b) The same curves, but, first, vertically shifted to coincide in the rightmost point, and then uniformly vertically offset to display the oscillations correlating with the QPC conductance.} 
\end{figure}

Fig. \ref{fig:fig2} shows the calculated $\Delta \Phi$ and $G$ dependences on $V_\mathrm{G}$. Different curves correspond to different transverse dimensions of the QPC $t$. Opening of the QPC leads to a decrease in the resonant frequency saturated at large $V_\mathrm{G}$ values corresponding to the regime of pure electrostatic screening. An increase in $t$ results in an increase in the absolute $\Delta\Phi$ values. The transition between the closed and opened states of the QPC is accompanied by oscillations of the resonant frequency (see Fig. \ref{fig:fig2} (b)). Local minima of the resonant frequency are observed at the transitions between $G$ plateaus, similarly to our experiment. The amplitude of the oscillations rapidly decreases with increasing subband index, likewise in our experiment the second oscillation is weaker than the first one. Thus, the experimentally observed oscillations may originate from the oscillatory ability of the QPC to screen the piezoelectric charge.

The fact that the third and the fourth of the measured oscillations (see Fig. \ref{fig:fig1} (d)) are deeper than the first two, unlike in the calculated curves displayed in Fig. \ref{fig:fig2} (b), can be explained by the following reasoning. Each of the curves in Fig. \ref{fig:fig2} corresponds to a fixed $V_\mathrm{G}$-independent QPC transverse size $t$. However, an abrupt change in $t$ is expected each time a new subband starts filling \cite{Drexler_1994_qcapacitance}, because the spreading of the transverse wavefunctions increases with increasing subband index. Since the inter-subband energy spacings decrease with $V_\mathrm{G}$ increase in the experiment, we can conclude that the QPC confinement potential is sub-parabolic, and the wavefunction widening is more prominent for the top levels than for the bottom ones. Fig. \ref{fig:fig2} shows that a change in $t$, which could be modelled as inter-curve jumps, should lead to a prominent shift in the resonant frequency. Thus, the third and the fourth of the observed oscillations can be explained by an abrupt change of the system geometry accompanying the subband filling.

Provided that the observed effect is enhanced (by, for example, increasing the QPC length), it may be possibly utilized for studying the DOS features in quasi-1D systems caused by many-body phenomena.

To conclude, in a nanoelectromechanical system comprising a mechanical resonator coupled to a quantum point contact (QPC), the electronic state of the QPC influences the mechanical oscillations, leading to measurable changes in their resonant frequency. The frequency displays oscillations each time a new quantization subband starts filling, correlating with the QPC conductance. The observed oscillations can be explained by the piezoelectric effect and by the variations in the QPC ability to screen the piezoelectric fields induced by mechanical vibrations. Such variations can be expected because of both oscillatory density of states and abrupt changes of the quantum point contact geometry at the inter-plateau transitions. The discovered back action of quantum transport on mechanical motion opens up new prospects for studying the density of states, compressibility and quantum capacitance in quasi-one-dimensional electron systems.

\begin{acknowledgments}
The work is supported by Russian Science
Foundation (grant \textnumero22-12-00343 – experimental measurements and theoretical simulation) and the Ministry
of Science and Higher Education of The Russian Federation (project \textnumero FWGW-2022-0011 – characterization
of the initial structures).
\end{acknowledgments}

\appendix*
\section{Appendix: Calculation details}

For a quasi-1D system, the DOS per unit length is $D(\mu)=\frac{\sqrt{2m}}{\pi\hbar}\sum_i\frac{1}{\sqrt{\mu-E_i}}$, where $\mu$ is the chemical potential. To eliminate the need to calculate the electron density distribution in the transverse direction, we reduce the geometry of the QPC, together with its source and drain, to a thin cylinder of diameter $t$ (varied in the calculations) and of infinite length directed along $x$ axis (the radial coordinate is $r$), with $\phi$ displaying a hill at $x=0$, and electrochemical potential $\mu-e\phi$ being constant. The complex 3D geometry of the rest of the system is also reduced to the axially symmetric one, and we assume that the QPC cylinder is embedded into a thicker one with dielectric constant $\varepsilon=12$ and diameter $h=166$ nm (the resonator thickness). The thermally-smoothed DOS can be calculated as
\begin{equation}
\tilde{D}(x)=\int\displaylimits_{-\infty}^{\infty}D(E)\frac{\partial F(\frac{E-\mu(x)}{kT})}{\partial \mu}\mathrm{d}E,
\end{equation}
where $F(\zeta)=\left(1+\exp\zeta\right)^{-1}$ is the Fermi function.

The potential hill at $x=0$ (the saddle of the QPC) is created by applying boundary condition $\phi(x)=V_0+V_\mathrm{G}\exp{\left(-x^2/L_\mathrm{G}^2\right)}$
at $r=R$, where $V_0>0$ is chosen in such a way that $\mu$ is close to that of 2DEG at a large $x$, and $V_\mathrm{G}<0$ is the varied gate voltage. Once the problem is solved, conductance of the QPC can be evaluated from $\mu$ value at $x=0$ \cite{Buttiker_1990}. Parameter $R=1.15h$ is chosen in such a way that the period of the conductance plateaus with respect to $V_\mathrm{G}$ is close to the experimentally measured one. $L_\mathrm{G}=0.25h$ in the calculations.

Bending of a thin piezoelectric membrane induces a volume bound charge, uniform along its thickness, compensated by the opposite surface charge \cite{Shevyrin_2016_piezo}. In our axially-symmetric model, we introduce the piezoelectric charge in a similar way and assume that
\begin{eqnarray}
\rho_\mathrm{piezo}(\xi,r,x)=\rho_0(\xi)\left[\Theta\left(\frac{h}{2}-r\right)-\delta\left(r-\frac{h}{2}\right)\right]\\ \nonumber\times\exp\left(-\frac{x^2}{w^2}\right),
\end{eqnarray}
where $\Theta(r)$ is the Heaviside step function, and $\delta(r)$ is the delta-function. In $x$ direction, the charge is concentrated in a region of width $w$ centered at zero. A reasonable value of $w$ is the resonator width 1.3 $\mu$m at the clamped edge in our case.

The Thomas-Fermi model is expected to give reasonable results, provided that the electron wavelength is small in comparison to the characteristic spatial size of potential fluctuations. Thus, we do not expect correct results at the gate voltages close to the pinch-off, where the wavelength tends to infinity.

\providecommand{\noopsort}[1]{}\providecommand{\singleletter}[1]{#1}%

\end{document}